\documentclass[fleqn,10pt]{wlscirep}
\usepackage[utf8]{inputenc}
\usepackage[T1]{fontenc}

\usepackage{enumitem}

\title{Behavioural Movement Strategies in Cyclic Models}

\author[1, *]{B. Moura}
\author[1,2,**]{J. Menezes}

\affil[1]{Escola de Ci\^encias e Tecnologia, Universidade Federal
do Rio Grande do Norte\\
Caixa Postal 1524, 59072-970, Natal, RN, Brazil}
\affil[1]{Institute for Biodiversity and Ecosystem Dynamics,
University of Amsterdam, Science Park 904, 1098 XH Amsterdam, The
Netherlands}

\affil[*]{bianpmoura@gmail.com}
\affil[**]{jmenezes@ect.ufrn.br}

\begin{abstract}
The spatial segregation of species is fundamental to ecosystem formation and stability. Behavioural strategies may determine where species are located and how their interactions change the local environment arrangement.
In response to stimuli in the environment, individuals may move in a specific direction instead of walking randomly. This behaviour can be innate or learned from experience, and allow the individuals to conquer or the maintain territory, foraging or taking refuge. 
We study a generalisation of the spatial rock-paper-scissors model where individuals of one out of the species may perform directional movement tactics. Running a series of stochastic simulations, we investigate the effects of the behavioural tactics on the spatial pattern formation and the maintenance of the species diversity. We also explore a more realistic scenario, where not all individuals are conditioned to perform the behavioural strategy or have different levels of neighbourhood perception. Our outcomes show that self-preservation behaviour is more profitable in terms of territorial dominance,
with the best result being achieved when all individuals are conditioned and have a long-range vicinity perception. On the other hand, invading is more advantageous if part of individuals is conditioned and if they have short-range neighbourhood perception.
Finally, our findings reveal that the self-defence strategy is the least jeopardising to biodiversity which can help biologists to understand population dynamics in a setting where individuals may move strategically.
\end{abstract}

\begin{document}

\flushbottom
\maketitle

\section*{Introduction}
The understanding of the spatial segregation of species is crucial for ecology\cite{ecology}. 
It is well known that ecosystem formation and stability depend on the interaction among individuals \cite{Nature-bio, BUCHHOLZ2007401}. In this sense, it has been enlightening the outcomes from experiments with bacteria \textit{Escherichia coli} that revealed that space plays a vital role in preserving biodiversity \cite{Coli,bacteria}. It was observed that only cyclic dominance is not enough to maintain biodiversity, but individuals must interact locally, forming spatially detached domains\cite{Allelopathy}. 
The cyclic dominance among the bacteria strains is described by the rock-paper-scissors rules, where scissors cut paper, paper wraps rock, rock crushes scissors \cite {bacteria,Directional1, Directional2}. This type of spatial interaction between species has also been observed in groups of lizards \cite{lizards} and coral reefs\cite{Extra1}.  For this reason, stochastic simulations of the rock-paper-scissors game have been widely used to investigate biological systems\cite{Szolnoki_2020, Szolnoki-JRSI-11-0735}. In these models, random mobility competes with local interactions such as reproduction and selection, promoting (low mobility) or jeopardising (high mobility) biodiversity \cite{Reichenbach-N-448-1046}. 
Recently, some authors have dedicated attention to study population dynamics in spatial systems where individuals move according to behavioural strategies.
For example, individuals' dispersal rate may vary according to the resources available in their habitat.  In this case, it has been shown that species coexistence is strongly affected by the adaptive movement\cite{Korea}. Another approach considered that species react to the presence of others, developing directional movement\cite {Potts}. Using a telegrapher-taxis formalism, the authors analysed the differential equations and concluded that directional movement has a significant effect on the spatial distribution of the species. 

In this paper, we study cyclic nonhierarchical systems where individuals of one out of species move motivated by a stimulus in the environment
\cite{Motivation1,Dispersal,butterfly}. The movement strategy depends on the individuals' behaviour and aims to increase the species territorial dominance \cite{Causes,ClassificationMovement}.
Here, we assume two types of behaviours to describe systems where individuals respond to a stimulus either instinctively (innate behaviour) or based on the experience (conditioned behaviour) \cite{EthologySite}. 
First, as a foraging behaviour - in which natural resources are exploited - we define two movement strategies: i) Attack tactic: a directional movement that allows the individuals to go straight to areas mostly occupied by the individuals they dominate \cite{ForagingBehaviour,Motivation2,Motivation3,BehaviouralAgression};
ii) Anticipation tactic: a directional movement prompt by stalking target individuals, going to patches where their incoming is likely \cite{anticipationsweden,catching,retina,courses,capture,archer}. 
Second, as a defence behaviour - a reaction to a stimulus to prevent any damage - we define the Safeguard tactic: individuals move towards territories mostly occupied by individuals that give them protection \cite{refuge1,refuge2}.
We aim to understand how the behavioural movement strategies change the spatial patterns and, consequently,  the highest density zones of each species. Also, we address the effects of the tactics on the species coexistence\cite{Reichenbach-N-448-1046}.

To execute a behavioural movement strategy, an individual senses its neighbourhood, identifying the direction with more target individuals\cite{innatemite,Odour,MovementProfitable}. As this ability varies among species, we define a perception radius to describe how far the individual can perceive the vicinity\cite{Olfactory,perception}.
It has been suggested that some individuals do not perform behavioural strategies. This happens because they have not yet learned or, somehow, they cannot put the tactic into practice  \cite{doi:10.1002/ece3.4446,innatemite,SabelisII}. Therefore, to make the model more realistic, we describe the individual's ability to execute the tactic by defining a conditioning factor.  Our goal is to discover what perception radius and conditioning factor ensures the highest spatial density for the species whose individuals perform the movement tactics.

\begin{figure*}
	\centering
	\includegraphics[width=100mm]{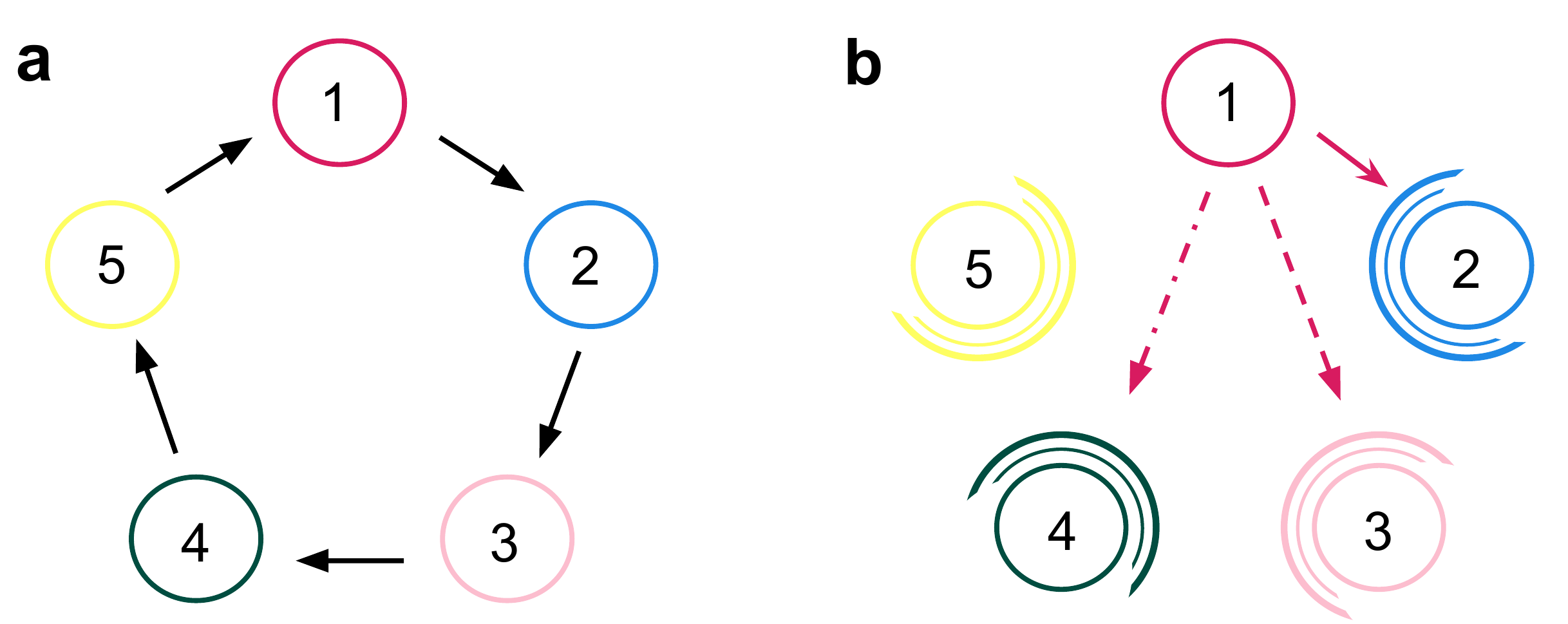}
    \caption{Selection rules and directional movement tactics. ({\bf a}) Illustration of the selection rules among the species, which represents a generalisation of the rock-paper-scissors game. ({\bf b}) Illustration of the directional movement tactics for individuals of species $1$. The solid ruby line represents the Attack tactic, where individuals move towards the direction with more individuals of species $2$. The dashed ruby line shows the Anticipation tactic, that is a movement towards the path with more individuals of species $3$. The dashed-dotted ruby line illustrates how individuals move when they perform the Safeguard tactic, going towards the direction with more individuals of species $4$. The concentric circumference arcs in the right panel illustrate that individuals of species $2$, $3$, $4$, and $5$ always move randomly.}
  \label{fig1}
\end{figure*}

\begin{figure*}[h]
	\centering
	\includegraphics[width=175mm]{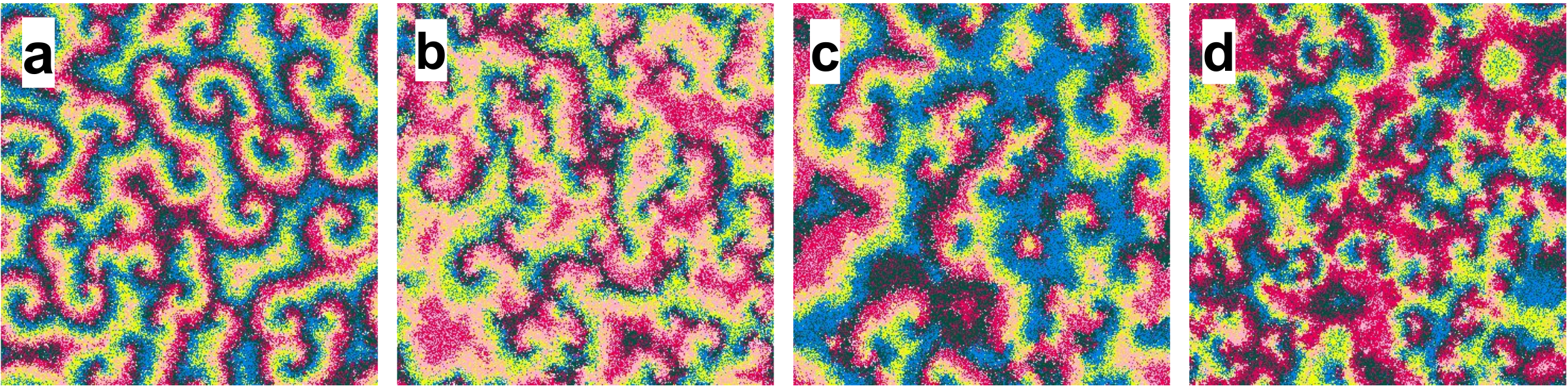}
    \caption{Snapshots of simulations of the generalisation of the rock-paper-scissors game illustrated in Fig. \ref{fig1} running in square lattices with $500^2$ grid points. Each dot shows either an individual (according to the colour scheme in Fig. \ref{fig1}) or an empty site (white dot). All simulations started from the same random initial conditions. The snapshots show the spatial patterns for the standard model ({\bf a}), Attack ({\bf b}), Anticipation ({\bf c}), and Safeguard tactics({\bf d}), respectively, at $t\,=\,5000$ generations. See also the videos for the whole simulation for the Standard case (https://youtu.be/Ndvk6Rg57m4), Attack (https://youtu.be/JGhkDAHSo74), Anticipation (https://youtu.be/ZZp9QlOfv2Q), and Safeguard (https://youtu.be/eFxWdLhIOuQ). The results were obtained for $R=3$ and $s\,=\,r\,=\,m\,=\,1/3$.}
  \label{fig2}
\end{figure*}

\section*{Results}

\begin{figure*}
	\centering
\includegraphics[width=155mm]{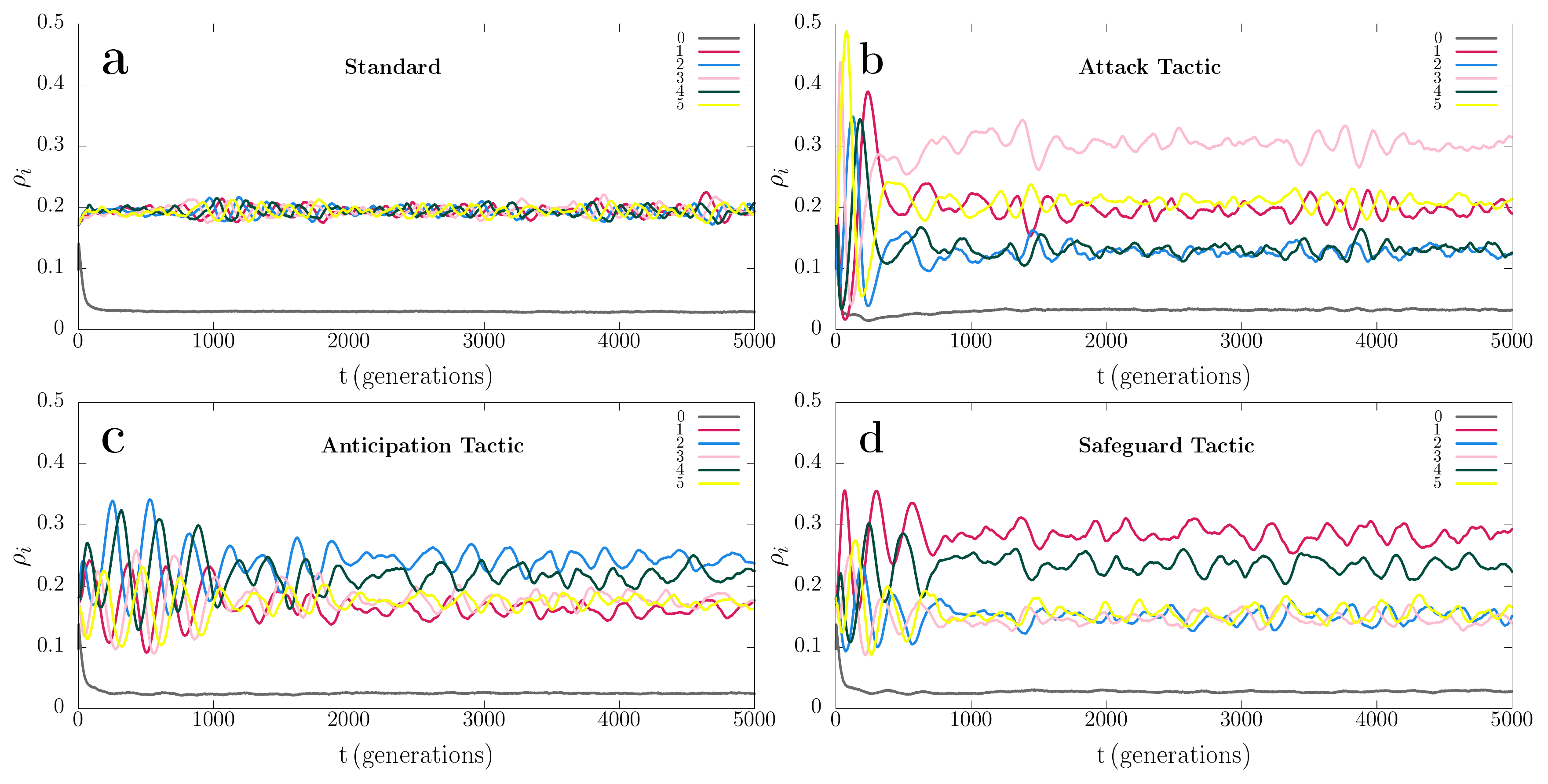}
    \caption{Temporal changes of spatial densities $\rho_i$, with $i=0,1,2,3,4,5$,  where $i=0$ indicates the empty spaces and $i=1,2,3,4,5$ represent the species according to the illustration in Fig.~\ref{fig1}. The results were obtained from the simulations presented in Fig.~\ref{fig2}.  ({\bf a}) Standard case, https://youtu.be/Ndvk6Rg57m4. ({\bf b}) Attack tactic, https://youtu.be/JGhkDAHSo74. ({\bf c}) Anticipation tactic, https://youtu.be/ZZp9QlOfv2Q. ({\bf d}) Safeguard tactic, https://youtu.be/eFxWdLhIOuQ.}
  \label{fig3}
\end{figure*}

\begin{figure*}
	\centering
		\includegraphics[width=155mm]{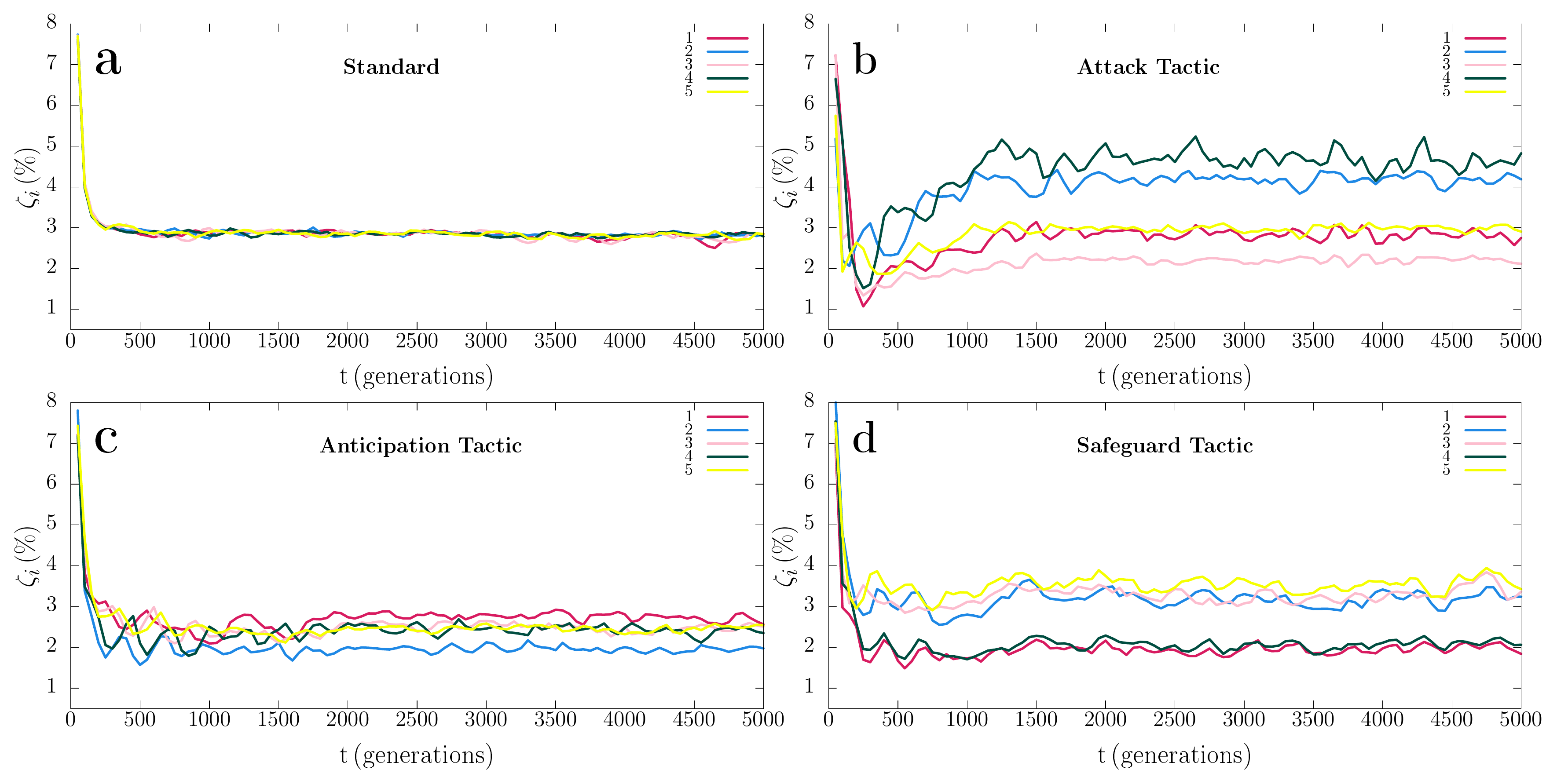}
    \caption{Dynamics of the species selection risks $\zeta_i$ for the simulations presented in Fig.~\ref{fig2}. The colours indicate the species following the scheme in Fig.~\ref{fig1}. ({\bf a}) Standard case, https://youtu.be/Ndvk6Rg57m4. ({\bf b}) Attack tactic, https://youtu.be/JGhkDAHSo74. ({\bf c}) Anticipation tactic, https://youtu.be/ZZp9QlOfv2Q. ({\bf d}) Safeguard tactic, https://youtu.be/eFxWdLhIOuQ.}
  \label{fig4}
\end{figure*}

We first investigated how directional movement tactics affect spatial patterns (see Methods). Figure ~\ref{fig2} shows the spatial patterns obtained from a $500^2$ simulation running for a timespan of $5000$ generations. Figure~\ref{fig2}\textcolor{blue}{(a)}, \ref{fig2}\textcolor{blue}{(b)}, \ref{fig2}\textcolor{blue}{(c)}, and \ref{fig2}\textcolor{blue}{(d)} show the spatial patterns captured at $t=5000$ for the standard model, Attack, Anticipation, and Safeguard directional movement tactics, respectively. See also the videos for the entire simulation for the standard case (https://youtu.be/Ndvk6Rg57m4), Attack (https://youtu.be/JGhkDAHSo74), Anticipation (https://youtu.be/ZZp9QlOfv2Q), and Safeguard (https://youtu.be/eFxWdLhIOuQ). 
The colours follow the scheme in Fig. ~\ref{fig1}, where ruby, blue, pink, green, and yellow dots represent individuals of species $1$, $2$, $3$, $4$, and $5$, respectively. White dots indicate empty spaces.
Figure~\ref{fig3}\textcolor{blue}{(a)} depicts the results for the dynamics of the species densities for the standard model whereas Figs.~\ref{fig3}\textcolor{blue}{(b)}, \ref{fig3}\textcolor{blue}{(c)}, and \ref{fig3}\textcolor{blue}{(d)} show how $\rho_i$ changes when individuals of species $1$ use Attack, Anticipation, and Safeguard tactics respectively. Additionally, in Fig.~\ref{fig4}, the effects of the behaviour on the selection risks of each species were computed by $\zeta_i$, that shows the probability of one individual of species $i$ disappearing within a unit time interval. Figure~\ref{fig4}\textcolor{blue}{(a)} depicts the case where species $1$ move randomly, while Figs.~\ref{fig4}\textcolor{blue}{(b)}, \ref{fig4}\textcolor{blue}{(c)}, and \ref{fig4}\textcolor{blue}{(d)} show the selection risks if individuals of species $1$ move according to the Attack, Anticipation, and Safeguard tactics respectively. In Figs.~\ref{fig3} and \ref{fig4}, the colours identify the species according to Fig. \ref{fig1}, whereas in Fig.~\ref{fig3} the grey line shows the density of empty spaces.

Initially, individuals of all species are distributed aleatorily on the grid. Because of the random initial condition, selection interactions are frequent in the initial stage of the simulation. The result is the formation of spirals whose adjacent arms are mostly occupied by individuals of species that do not select each other. In the standard model, the spirals are symmetric (Fig. \ref{fig2}\textcolor{blue}{(a)}), leading to a cyclic territorial dominance of the species (Fig.~\ref{fig3}\textcolor{blue}{(a)}) and selection risks (Fig.~\ref{fig4}\textcolor{blue}{(a)}). 
This symmetry is broken if individuals of species $1$ perform a directional movement tactic. 
Firstly, if the individuals of species $1$ use the Attack tactic, they have more chances of selecting because they move towards the direction
with more individuals of species $2$ - even though the selection probability $s$ is the same for all species. The higher selection rate for species $1$ is responsible for the alternating territorial dominance verified in the early stage of the pattern formation shown in the video https://youtu.be/JGhkDAHSo74 (see Ref.~\cite{uneven}).  The reason is that when the number of individuals of species $2$ decreases, the population of species $3$ rises, reducing the population of species $4$ and allowing the population growth of species $5$. The consequence is that more individuals of species $5$ lead to a higher selection risk of species $1$, as it is showed in Fig.~\ref{fig4}\textcolor{blue}{(b)}. Therefore, although the higher selection rate for species $1$, 
it does not dominate when individuals perform the Attack tactic. Instead, species $3$ is more abundant, as it is depicted in Fig.~\ref{fig3}\textcolor{blue}{(b)}.
Secondly, if the individuals of species $1$ use the Anticipation tactic (they go towards the direction with more individuals of species $3$), species $2$ is preserved, and its population grows. There is a consequent reduction of the number of individuals of species $3$, allowing the population of species $4$ to grow, which limits the number of individuals of species $5$. Even though this scenario appears to be favourable to species $1$, the fewer individuals of species $5$ do not imply a less selection risk for species $1$, as it is shown in Fig.~\ref{fig4}\textcolor{blue}{(c)}. The selection risk of species $1$ is high because its population growth is restricted since individuals of species $1$ go apart from individuals of species $2$, making it difficult to conquer territory. 
Thirdly, if the Safeguard tactic is used, the population growth of species $1$ is expected due to the protection provided by individuals of species $4$ against eventual attacks of individuals of species $5$. Mostly, when the individuals of species $5$ approach individuals of species $1$, they find guards, which destroy them. This effect is reinforced with the population growth of species $2$, which controls the population size of species $3$, leading to a higher abundance of individuals of species $4$ - the more individuals of species $4$, the more available refuges for species $1$. The result is a relevant decreasing in the selection risk of the species $1$, as it is depicted in Fig.~\ref{fig4}\textcolor{blue}{(d)}. As a consequence, the density of species $1$ is the highest, according to Fig. \ref{fig3}, that shows that species $1$ dominate during the entire simulation, with species $4$ being the second most abundant one. 

\subsection*{Autocorrelation Function}
\begin{figure*}
	\centering
	\includegraphics[width=160mm]{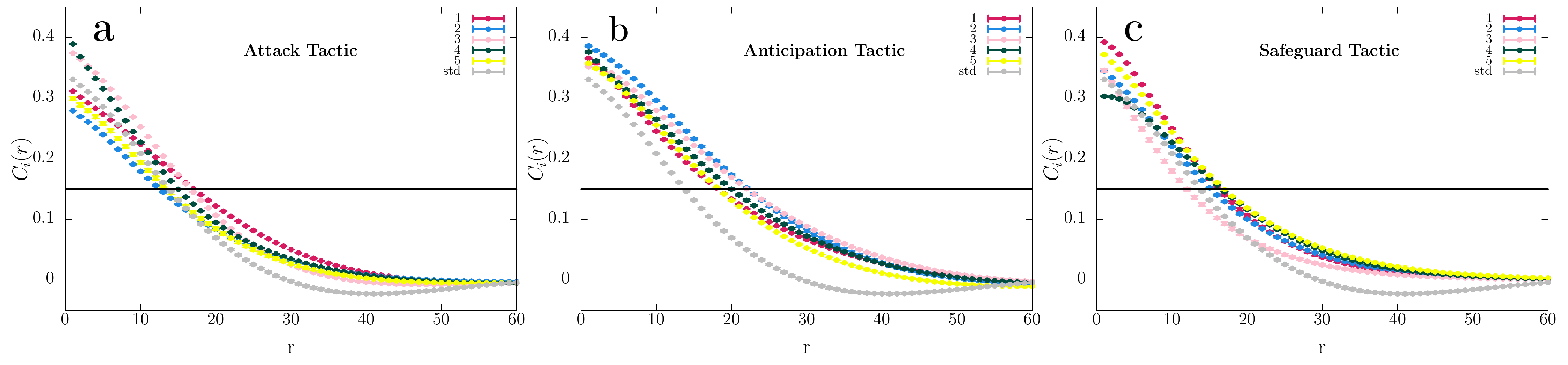}
    \caption{Autocorrelation functions $C_i$. The 
colours follow the scheme in Fig. \ref{fig1}. ({\bf a}), ({\bf b}) , and ({\bf c}) depict the cases where individuals of species $1$ use Attack, Anticipation, and Safeguard tactics, respectively. Grey circles show the results for the autocorrelation function for the standard model (std), that is the same for every species. The error bars indicate the standard deviation. The horizontal black line indicates the threshold assumed to calculate the characteristic length. The results were obtained using $R=3$ and $s\,=\,r\,=\,m\,=\,1/3$. }
  \label{fig5}
\end{figure*}

To quantify the effects of the directional movement tactics on the spatial patterns, we calculated the spatial autocorrelation function $C_i(r)$, with  $i=1,...,5$ (See Methods). 
Figures~\ref{fig5}\textcolor{blue}{(a)}, \ref{fig5}\textcolor{blue}{(b)}, and \ref{fig5}\textcolor{blue}{(c)} show the spatial autocorrelation function for the cases where individuals of species $1$ use the Attack, Anticipation, and Safeguard tactics, respectively.  The error bars indicate the standard deviation from a set of $100$ simulations running in square lattices of $500^2$ grid points. 
We computed the characteristic length $l_i$, defined as $C(l)=0.15$ in every case, as illustrated by the horizontal black line.
In the standard model, the characteristic length is the same for all species: $l_i = 14$, with $i=1,..,5$. However, if individuals of species $1$ moves according to the Attack tactic, the characteristic length of species $1$, $3$ , and $4$ enlarges ($l_1 = l_3 = 17$, and $l_4 = 15$) while the characteristic length of species $2$ and $5$ decreases ($l_2 = 12$ and $l_5 = 0.13$). For the Anticipation tactic, the characteristic length enlarges for all species ($l_1 = 18$, $l_2 = l_3 = 22$, $l_4 = 20$, and $l_5 = 18$). Finally, for the Safeguard tactic, with exception of the species $3$ whose characteristic length decreases ($l_3 = 12$), all other species have an elongation in the characteristic length ($l_1 = 16$, $l_2 = 15$, $l_4 = l_5 = 17$).

\subsection*{The Influence of the Perception Radius}

\begin{figure*}[h]
	\centering
		\includegraphics[width=165mm]{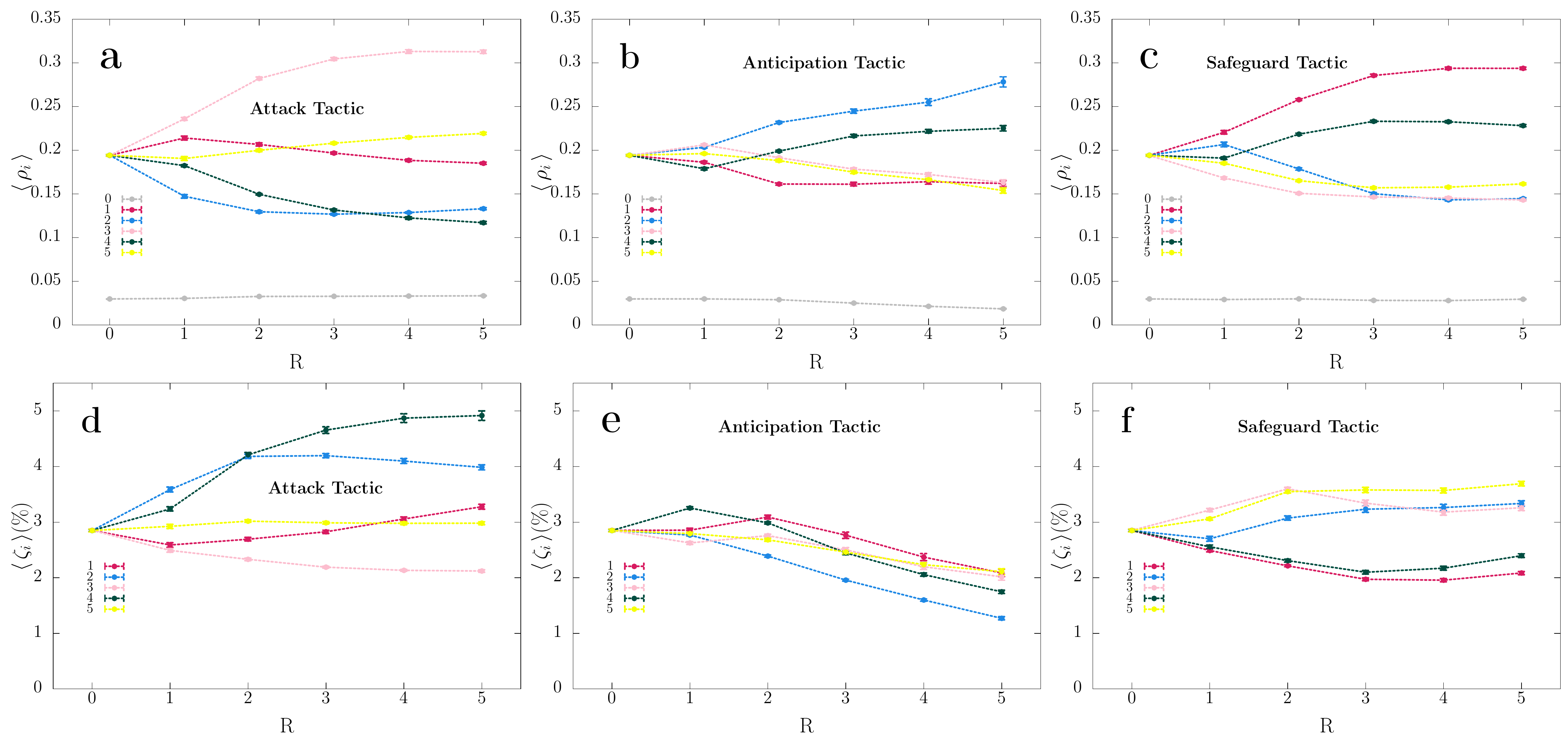}
    \caption{Mean species densities $\langle\rho_i\rangle$ and mean selection risks $\langle\zeta_i\rangle$ as a function of the perception radius $R$. $R=0$ represents the standard case.  ({\bf a}), ({\bf b}), and ({\bf c}) show $\langle\rho_i\rangle$ for Attack, Anticipation, and Safeguard tactics, respectively. ({\bf d}), ({\bf e}), and ({\bf f}) depict $\langle\zeta_i\rangle$ for Attack, Anticipation, and Safeguard tactics, respectively. The colours are the same as in Fig.~\ref{fig3} and ~\ref{fig4}. The error bars indicate the standard deviation. The results were obtained using $R=3$ and $s\,=\,r\,=\,m\,=\,1/3$.}
  \label{fig6}
\end{figure*}

To understand the role of the perception radius on the behavioural strategies, we run a series of simulations for $1\,\leq\,R\, \leq\,5$. Figures~\ref{fig6}\textcolor{blue}{(a)}, \ref{fig6}\textcolor{blue}{(b)}, and \ref{fig6}\textcolor{blue}{(c)} depict the mean spatial densities $\langle\rho_i\rangle$, with $i=0,...,5$, for the Attack, Anticipation, and Safeguard tactics used by individuals of species $1$, respectively ($\langle\rho_0\rangle$ indicates the density of empty spaces).
Figures \ref{fig6}\textcolor{blue}{(d)}, \ref{fig6}\textcolor{blue}{(e)}, and \ref{fig6}\textcolor{blue}{(f)} depict the selection risks $\langle\zeta_i\rangle$, with $i=1,...,5$, for the Attack, Anticipation, and Safeguard tactics, respectively. 
Each circle shows the mean value, while the error bars represent the standard deviation.  We also calculated the variation coefficient for our statistical results.

For the standard model, $\langle\rho_0\rangle \approx 0.03$, $\langle\rho_i\rangle \approx 0.194$, and  $\langle\zeta_i\rangle \approx 0.028$ for $i=1...,5$. In the Attack tactic, the directional mobility is advantageous for species $1$ for $R < 3$. Figure~\ref{fig6}\textcolor{blue}{(a)} shows that $\langle\rho_2\rangle\,<\, 0.194$ for any $R$. In comparison with the standard case, this indicates a harmful effect on the population of species $2$, benefitting species $3$. Indeed, the selection risk of species $2$ is always higher than in the standard model, as it is depicted in Fig.~\ref{fig6}\textcolor{blue}{(d)}. As the perception radius grows, the damage on the population of species $2$ becomes more significant. But, a higher $\langle \zeta_2 \rangle$ does not imply a growth of the population of species $1$. For the Anticipation tactic, for a large $R$, the chances of the direction with more individuals of species $3$ attracting individuals of species $1$ are greater, i.e., it is more likely that individuals of species $1$ discard the path with more individuals of species $2$. This is propitious to species $2$ conquer territory. 
For $R>2$, this territorial dominance is such significant that allows that individuals of species $1$ find individuals of species $2$ due to the Anticipation movement tactic. However, although the Anticipation tactic represents a profit in terms of spatial density for species $1$ for $R>2$, it is not advantageous compared to the standard mobility: $\langle\rho_1\rangle\,<\, 0.194$ for any perception radius. Furthermore, the Anticipation movement tactic by individuals of species $1$ provokes a reduction of selection risks for all species. This effect is reflected in a lower density of empty spaces, and becomes more relevant when $R$ enlarges. Finally, according to Fig. \ref{fig6}\textcolor{blue}{(c)}, the Safeguard tactic is profitable for species $1$ when compared with the standard model, irrespective of $R$. Moreover, the larger the perception radius, the more efficient the behavioural strategy is - reducing the selection risk of species $1$ becomes significant. However, for $R>3$, the high density of the species $1$ results in small unavoidable population growth of species $5$, which controls the population growth of species $1$. 

\subsection*{The Role of the Conditioning Factor}

\begin{figure*}[h]
	\centering
	\includegraphics[width=110mm]{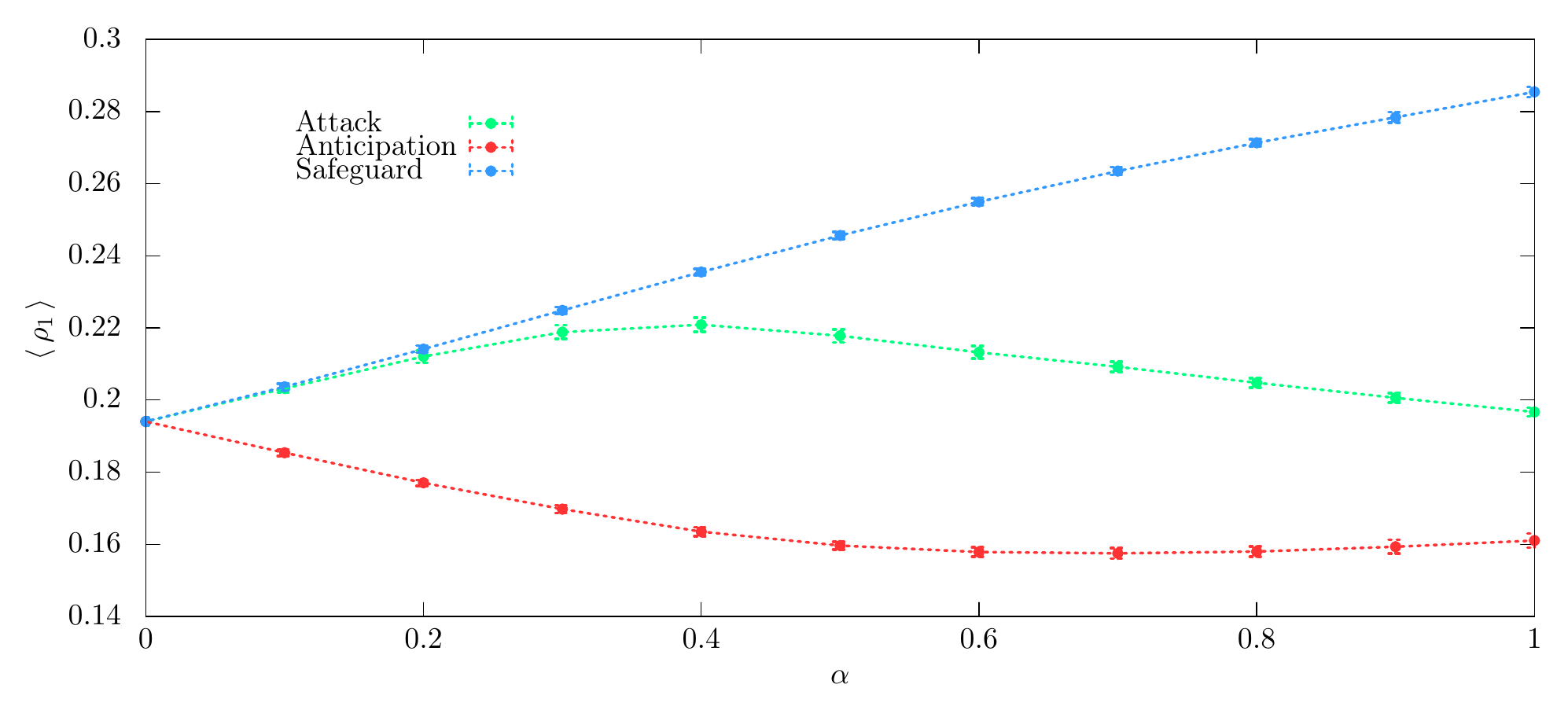}
    \caption{Mean density of species $1$ in terms of the conditioning factor $\alpha$. The green, red, and blue dashed lines show the results for the cases of individuals of species $1$ moving according to Attack, Anticipation, and Safeguard tactics, respectively. The error bars show the standard deviation. The results were obtained using $R=3$ and $s\,=\,r\,=\,m\,=\,1/3$.}
  \label{fig7}
\end{figure*}

We studied how the proportion of conditioned individuals of species $1$ influences the results. We calculated the average density of species $1$, $\langle\rho_1\rangle$ for the entire range of conditioning factor, $0\,\leq\,\alpha\,\leq 1$.  Figure \ref{fig7} depicts the variation of $\langle\rho_1\rangle$ in terms of $\alpha$, with $\alpha\,=\,0$ representing the standard case. The results show that Anticipation tactic is disadvantageous for species $1$: the spatial densities $\langle\rho_1\rangle$ is lower than the standard case, irrespective of $\alpha$. In contrast, Safeguard tactic is always advantageous - the more frequently the Safeguard tactic is used, the greater the fraction of the grid occupied by individuals of the species $1$ is. Concerning to the Attack tactic, the results show that for $0\,<\,\alpha\,\leq\,1$, there is a growth of $\langle\rho_1\rangle$, that is maximum for $\alpha\,=\,0.4$.  

\subsection*{Coexistence Probability}
\begin{figure*}
	\centering
	\includegraphics[width=110mm]{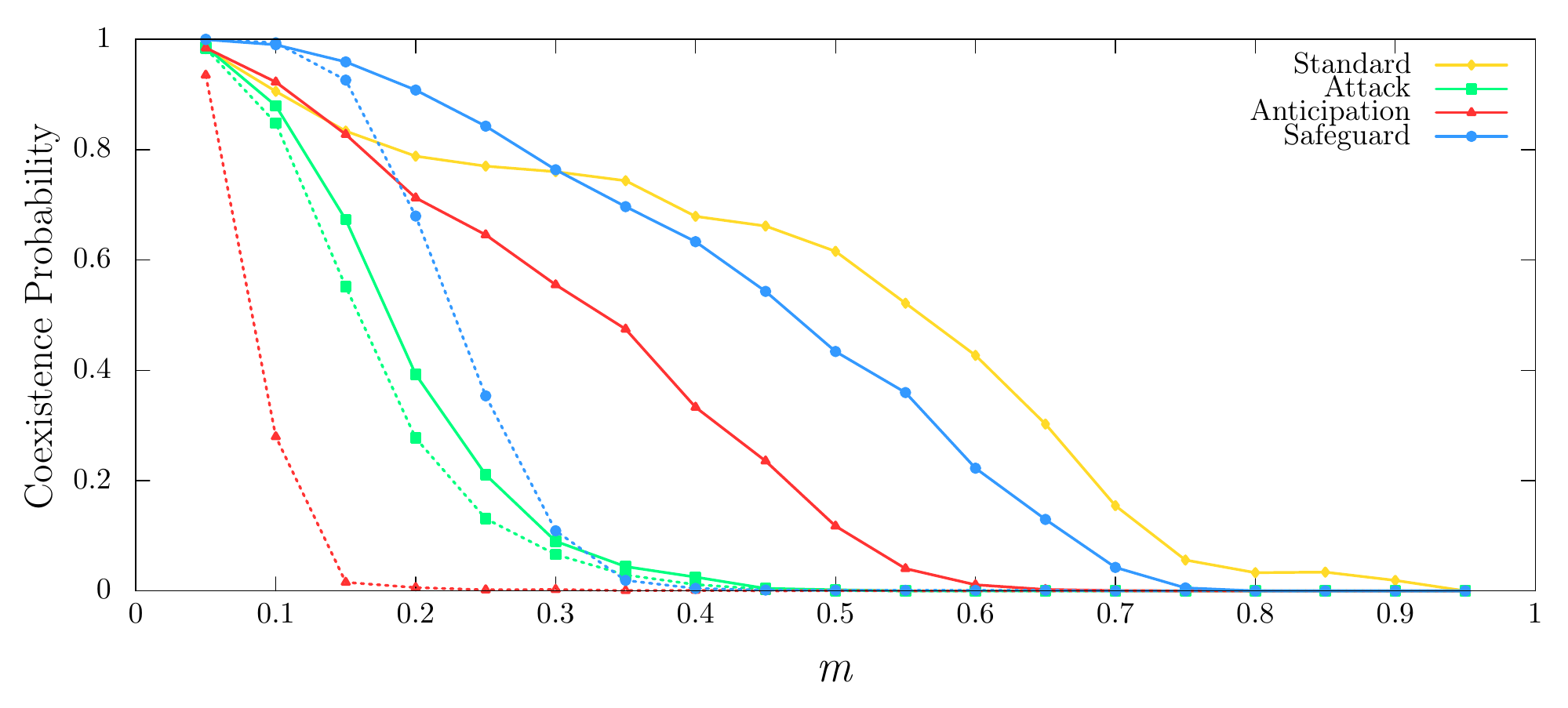}
 \caption{Coexistence probability as a function of the mobility probability $m$. The solid yellow line shows the coexistence probability for the standard model. The green, red, and blue lines show the results for the case of individuals of species $1$ moving according to Attack, Anticipation, and Safeguard directional tactics, respectively. Solid lines and dashed lines show the results for $R=2$ and $R=4$, respectively. The results were obtained by running $2000$ simulations in lattices with $100^2$ grid points running until $100^4$ generations, with $s\,=\,r\,=\,(1-m)/2$.}
  \label{fig8}
\end{figure*}

To observe the effects of the directional movement tactics on biodiversity, we calculated the coexistence probability. To this purpose, we performed $2000$ simulations using $100^2$ lattices, running until $10000$ generations, for a wide range of mobility probability $m$.
The yellow line in Figure \ref{fig8} shows the coexistence probability for the standard model. The green, red, and blue lines show the results when individuals of species $1$ use Attack, Anticipation and Safeguard directional tactics, respectively. Solid lines and dashed lines show the results for $R=2$ and $R=4$, respectively. Generally speaking, the chances of individuals of all species remain at the end of the simulation decrease as $m$ grows, independent of the specific behavioural tactic. 
Furthermore, whether individuals of species $1$ move directionally, the coexistence is less probable to maintain. 
The results indicate that for small $R$, Anticipation is the tactic that threatens the most coexistence, while for large $R$ is Attack tactic that most puts biodiversity in risk. In both cases, the Safeguard tactic is the directional movement that less endangers the coexistence.

\section*{Discussion}

We investigated a stochastic model of $5$ species, in which selection rules are a generalisation of the nonhierarchical cyclic rock-paper-scissors game. Based on the individual's behaviour, we explored various directional movement tactics for one out of the species.  Although we implemented directional mobility for only one species, all species are affected because of the cyclic selection rules.
The impact depends on the fraction of individuals that perform the behavioural strategy and how far they can perceive the neighbourhood.

To the best of our knowledge, this is the first time that simulations were performed assuming that all individuals within the perception range equally influence the directional movement tactic. Recently, some authors addressed a model for directional movement in which the individual's perception decreases exponentially with the distance\cite{PhysRevE.97.032415}. In that model, an individual chooses the direction to move mostly influenced by the first immediate neighbours, independent of its perception radius. Another difference in that model is that directional mobility is implemented for all species. This makes the effects of the individual's behaviour compensated by the cyclic game. Here, we focused on how a directional movement tactic used by one out of the species changes the spatial patterns and, consequently,  species spatial densities. We also investigate the impact on the coexistence probability.

The main result of our investigation indicates that the individual's behaviour plays a central role in population dynamics of spatial biological systems. Our statistical results are robust and reveal that the behaviour of self-preservation is more profitable in terms of population growth. The simulations showed that Safeguard is the directional movement tactic that brings more profit in terms of spatial density when compared with the standard model. More, the highest gain is achieved if: i) the perception radius of the individuals is maximum; ii) the totality of individuals always perform the Safeguard tactics. In opposite, the Attack tactic is more beneficial for the species if individuals have a short perception radius and intercalate between directional and random motion (for $R=3$, the higher spatial density is achieved when only $40\%$ of individuals move directionally). Finally, the results suggest that Anticipation tactic is disadvantageous, independent of the perception radius and the fraction of the conditioned individuals. This outcome may contribute for explaining the persistence of species whose part of the individuals are not able to learn or perceive the neighbourhood (see \cite{innatemite,SabelisII}, for example). 

The results can be extended to systems with a generic number of species $N$, with odd $N$, where spirals with $N$ arms form the spatial patterns. The adjacent arms are mostly occupied by individuals of species that do not select each other. For example, for a system with $N=5$ species, the spiral arms are occupied by individuals of species in the following order: $\{i+1; i-2; i; i+2; i-1\}$. For an arbitrary $N$, there are $(N-3)/2$ arms between the spatial concentrations of individuals of species $i$ and $i+1$, with $i=1,..,N$. This means that for the Attack tactic to be efficient, the perception radius $R$ must be larger as the number of species increases. In contrast, regardless of the number of species, Anticipation and Safeguard tactics efficiency is not ruined. The reason is that the spiral arms occupied mostly by individuals of the species $i$ are always adjacent to the arms populated by individuals of the species $i-2$ (outer arm) and to the arms mainly formed by individuals of the species $i+2$ (inner arm).  In the specific case of Anticipation tactic, individuals of species $i$ discover where the individuals $i+2$ concentrate, moving towards the opposite direction to the spiral wave propagation. This reduces the selection risks of all species, and, consequently, the density of empty spaces. On the other hand, if the Safeguard tactic is applied, groups of individuals of the $i-2$ species are perceived, making individuals of species $i$ to move towards the spiral wave propagation direction.

In our model, a single species evolves into one movement tactic, while individuals of the other species move randomly. In this scenario, when an individual of the species $i$ anticipates, it aims to arrive earlier in the areas where individuals of the species $i+1$ will multiply. However, individuals of species $i+1$ do not walk directly towards the place the individual of species $i$ is waiting for them, but it depends on the simulation stochasticity. Suppose that except for the species $i$, all other species use Attack tactic. Now, individuals of the species $i$ are guaranteed that individuals of the species $i+1$ will go wherever they are, intensifying the effects of the Anticipation tactics presented in this paper.  Likewise, let us consider that individuals of species $i$ use the Safeguard tactic, and every species else performs the Attack tactic. 
In this scenario, as individuals of species $i-1$ chase individuals of species $i$, the shelter offered by individuals of the species $i-2$ becomes more relevant. We conclude that Anticipation and Safeguard tactics may give more advantage in terms of population growth for one species if individuals of the others perform the Attack tactics. However, this effect may compromise biodiversity because if individuals of one out of the species move directionally, the coexistence probability decreases. Our findings may be useful to understand population dynamics and biodiversity and describe complex systems in other areas of nonlinear science. 
\section*{Methods}


In this work, we performed stochastic simulations of a cyclic nonhierarchical system composed of $5$ species. To this purpose, we implemented a standard numerical algorithm largely used to study spatial biological systems \cite{Reichenbach-N-448-1046,uneven,Szolnoki_2020}. We considered a generalisation of the rock-paper-scissors game for $5$ species, whose rules are illustrated in Fig.~\ref{fig1}\textcolor{blue}{(a)}. The arrows indicate a cyclic dominance among the species. Accordingly, individuals of species $i$ beat individuals of species $i+1$, with $i=1,2,3,4,5$.

The dynamics of individuals' spatial organisation occurs in a square lattice with periodic boundary conditions, following the rules: selection, reproduction, and mobility. We assumed the May-Leonard implementation so that the total number of individuals is not conserved \cite{leonard}. Each grid point contains at most one individual, which means that the maximum number of individuals is $\mathcal{N}$, the total number of grid points. 

Initially, the number of individuals is the same for all species, i.e., $I_i\,=\,\mathcal{N}/5$, with $i=1,2,3,4,5$ (there are no empty spaces in the initial state).  We prepared the initial conditions by distributing each individual at a random grid point. At each timestep, one interaction occurs, changing the spatial configuration of individuals.
The possible interactions are:
\begin{itemize}
\item 
Selection: $ i\ j \to i\ \otimes\,$, with $ j = i+1$, where $\otimes$ means an empty space; every time one selection interaction occurs, the grid point occupied by the individual of species $i+1$ vanishes.
\item
Reproduction: $ i\ \otimes \to i\ i\,$; when one reproduction is realised an individual of species $i$ fills the empty space.
\item 
Mobility: $ i\ \odot \to \odot\ i\,$, where $\odot$ means either an individual of any species or an empty site; an individual of species $i$ switches positions with another individual of any species or with an empty space.
\end{itemize}

In our stochastic simulations, selection, reproduction, and mobilities interactions occur with the following probabilities: $s$, $r$ and $m$, respectively. We assumed that individuals of all species have the same probabilities of selecting, reproducing and moving. The interactions were implemented by assuming the von Neumann neighbourhood, i.e., individuals may interact with one of their four nearest neighbours. The simulation algorithm follows three steps: i) sorting an active individual; ii) raffling one interaction to be executed; iii) drawing one of the four nearest neighbours to suffer the sorted interaction (the only exception is the directional mobility, where the neighbour is chosen according to the movement tactic). If the interaction is executed, one timestep is counted. Otherwise, the three steps are redone. Our time unit is called generation, defined as the necessary time to $\mathcal{N}$ timesteps to occur.

In our model, individuals of one out of the species can move into the direction with more individuals of a target species. The choice is based on the strategy assumed by species. We assumed three sorts of directional movement tactics:  
\begin{itemize}
\item
Attack tactic: an individual of species $i$ moves into the direction with more individuals of species $i+1$;
\item
Anticipation tactic: an individual of species $i$ goes towards the direction with a larger number of individuals of species $i+2$;
\item 
Safeguard tactic: an individual of species $i$ walk into the direction with a larger concentration of individuals of species $i-2$.
\end{itemize}
In the standard model, individuals of all species move randomly. 

We considered that only individuals of species $1$ perform the directional movement tactics, as illustrated in 
Figure \ref{fig1}\textcolor{blue}{(b)}.
The solid, dashed, and dashed-dotted lines represent the Attack, Anticipation, and Safeguard tactics, respectively.
The concentric circumference arcs show that individuals of species $2$, $3$, $4$, and $5$ always move randomly.
For implementing a directional movement, the algorithm follows the steps: i) it is assumed a disc of radius $R$ (the perception radius), in the active individual's neighbourhood;
ii) it is defined four circular sectors in the directions of the four nearest neighbours; iii) according to the movement tactic, the target species is defined: species $2$, $3$, and $4$, for Attack, Anticipation, and Safeguard tactics, respectively; iv) it is counted the number of individuals of the target species within each circular sector. Individuals on the borders are assumed to be part of both circular sectors; v) the circular sector that contains more individuals of the target species is chosen. In the event of a tie, a draw between the tied directions is made; vi) the active individual switches positions with the immediate neighbour in the chosen direction. The swap is also executed in case of the neighbour grid point is empty.

To observe the spatial patterns, we first performed a single simulation for the standard model, Attack, Anticipation, and Safeguard tactics. The realisations run in square lattices with $500^2$ grid points, for a timespan of $5000$ generations.  We captured $500$ snapshots of the lattice (in intervals of $10$ generations), that were used to make the videos of the dynamics of the spatial patterns showed in https://youtu.be/Ndvk6Rg57m4 (standard), https://youtu.be/JGhkDAHSo74 (Attack), https://youtu.be/ZZp9QlOfv2Q (Anticipation), and https://youtu.be/eFxWdLhIOuQ (Safeguard). The final snapshots were depicted in Fig. \ref{fig2}\textcolor{blue}{(a)}, Fig. \ref{fig2}\textcolor{blue}{(b)}, Fig. \ref{fig2}\textcolor{blue}{(c)}, and Fig. \ref{fig2}\textcolor{blue}{(d)}. Individuals of species $1$, $2$, $3$, $4$, and $5$ are identified with the colours ruby, blue, pink, green, and yellow, respectively; while white dots represent empty spaces.
The simulations were performed assuming selection, reproduction, and mobility probabilities: $s = r = m = 1/3$. The perception radius was assumed to be $R=3$.

The population dynamics were studied by means of the spatial density $\rho_i$, defined as the fraction of the grid occupied by individuals of species $i$ at time $t$, i.e., $\rho_i = I_i/\mathcal{N}$,  where $i=0$ stands for empty spaces and $i=1,...,5$ represent the species $1$, $2$, $3$, $4$, and $5$. The temporal changes in spatial densities of the simulations showed in Fig. \ref{fig2} were depicted in Fig. \ref{fig3}, where the grey, ruby, blue, pink, green, and yellow lines represent the densities of empty spaces and species $1$, $2$, $3$, $4$, and $5$, respectively.
We also computed how the selection risk of individuals of species $i$ changes in time. To this purpose, the algorithm counts the total number of individuals of species $i$ at the beginning of each generation. It is then counted the number of times that individuals of species $i$ are killed during the generation. The ratio between the number of selected individuals and the initial amount is defined as the selection risk of species $i$, $\zeta_i$. The results were averaged for every $50$ generations. Figure \ref{fig4} shows $\zeta_i\,(\%)$ as a function of the time for the simulations presented in Fig. \ref{fig2}. The ruby, blue, pink, green, and yellow lines show the selection risks of individuals of species $1$, $2$, $3$, $4$, and $5$, respectively.

To quantify the spatial organisation of the species, we studied the spatial autocorrelation function. This quantity measures how individuals of a same species are spatially correlated, indicating spatial domain sizes.  Following the procedure carried out in literature \cite{acf1,acf2,acf3,uneven,PhysRevE.97.032415}, we first calculated the Fourier transform of the spectral density as 
$C(\vec{r}') = \mathcal{F}^{-1}\{S(\vec{k})\}/C(0)$,
where the spectral density $S(\vec{k})$ is given by
$S(\vec{k}) = \sum_{k_x, k_y}\,\varphi(\vec{\kappa})$,
with $\varphi(\vec{\kappa}) = \mathcal{F}\,\{\phi(\vec{r})-\langle\phi\rangle\}$. The function $\phi(\vec{r})$ represents the species in the position $\vec{r}$ in the lattice (we assumed $0$, $1$, $2$, $3$, $4$, and $5$, for empty sites, and individuals of species $1$, $2$, $3$, $4$, and $5$, respectively). We then computed the spatial autocorrelation function as
$$C(r') = \sum_{|\vec{r}'|=x+y} \frac{C(\vec{r}')}{min (2N-(x+y+1), (x+y+1)}.$$
Subsequently, we found the scale of the spatial domains of species $i$, defined for $C(l_i)=0.15$, where $l_i$ is the characteristic length for species $i$.

We calculated the autocorrelation function by running $100$
simulations using lattices with $500^2$ grid points, assuming $s = r = m = 1/3$ and $R=3$. Each simulation started from different random initial conditions. We then captured each species spatial configuration after $5000$ generations to calculate the
autocorrelation functions. Finally, we averaged the autocorrelation function in terms of the radial coordinate $r$ and calculated the characteristic length for each species. We also calculated the standard deviation for the autocorrelation functions and the characteristic lengths. Figure~\ref{fig4} shows the comparison of the results for Attack, Anticipation, and Safeguard strategies with the standard model. The ruby, blue, pink, green, and yellow circles indicate the mean values for species $1$, $2$, $3$, $4$, and $5$, respectively. In the case of standard model, the mean values are represented by grey circles, which are the same for all species. The error bars show that standard deviation. The horizontal black line represents $C(l_i)\, =\, 0.15$.

To further explore the numerical results, we studied how the perception radius $R$ influences species spatial densities and selection risks. We calculated the mean value of the spatial species densities, $\langle\, \rho_i\,\rangle$ and the mean value of selection risks, $\langle\, \zeta_i\,\rangle$ from a set of $100$ simulations in lattices with $500^2$ grid points, starting from different initial conditions for $R=1,2,3,4,5$. We used $s\,=r\,=\,m\,=1/3$ and a timespan of $t=5000$ generations. The mean values and standard deviation were calculated using the second half of the simulations, thus eliminating the density fluctuations inherent in the pattern formation process. The results were shown in Fig.~\ref{fig6}, where the circles represent the mean values and error bars indicate the standard deviation. The colours are the same as in Fig.~\ref{fig3} and Fig.~\ref{fig4}. Furthermore, to verify the precision of the statistical results, we calculated the variation coefficient - the ratio between the standard deviation and the mean value. 

We studied a more realistic scenario where not all individuals of species $1$ can perform the directional movement tactics. For this reason, we defined the conditioning factor $\alpha$, with $0\,\leq\,\alpha\,\leq\,1$, representing the proportion of individuals of species $1$ that moves directionally. For $\alpha=0$ all individuals move randomly while for $\alpha=1$ all individuals move directionally.  This means that every time an individual of species $1$ is sorted to move, 
there is a probability $\alpha$ of the algorithm implementing the directional movement tactic, instead of randomly choosing one of its four immediate neighbours to switch positions. 
To understand the effects of the conditioning factor, we observed how the density of species $1$ changes for the entire range of $\alpha$, with intervals of $\Delta \alpha = 0.1$. The simulations were implemented for $R=3$ and $s\,=r\,=\,m\,=1/3$.
It was computed the mean value of the spatial density of species $1$,  $\langle\, \rho_1\,\rangle$, and its standard deviation from a set of $100$ different random initial conditions. The results were depicted in Fig.~\ref{fig7}, where the green, red, and blue dashed lines show $\langle\, \rho_1\,\rangle$ as a function of $\alpha$. The error bars indicate the standard deviation. 

Finally, we aimed to investigate how the directional movement tactics jeopardise species coexistence for a wide mobility probability range. 
Because of this, we run $2000$ simulations in lattices with $100^2$ grid points for $ 0.05\,<\,m\,<\,0.95$ in intervals of $ \Delta\, m\, =\,0.05$, with $R=2$ and $R=4$. The simulations started from different random initial conditions and run for a timespan of $10000$ generations. 
Coexistence happens if at least one individual of all species is present at the end of the simulation, $I_i (t=5000) \neq 0$ with $i=1,2,3,4,5$. Otherwise, the simulation results in extinction. 
The coexistence probability is the fraction of implementations which results in coexistence.
The simulations were performed for two values of perception radius, $R=2$ and $R=4$; the selection and reproduction probabilities were assumed to be $s\,=\,r\,=\,(1-m)/2$. The results were depicted in Fig.~\ref{fig8}, where yellow, green, red, and blue lines show the coexistence probability as a function of $m$ for the standard model, Attack, Anticipation, and Safeguard tactics, respectively.
The solid and dashed lines show the results for $R=2$ and $R=4$, respectively.

\bibliography{ref}

\section*{Acknowledgements}

We thank CNPq, FAPERN, ECT/UFRN, and IBED/University of Amsterdam for financial and computational support.

\section*{Author contributions statement}
J. M. conceived the research project. B.M. performed the numerical simulations. J. M. and B.M. analysed the results, wrote and revised the manuscript.
\end{document}